\documentclass[a4paper,11pt]{article}
\usepackage{pos}

\newcommand{\zmax}{z_{\textrm{max}}}

\let\oldbibliography\thebibliography
\renewcommand{\thebibliography}[1]{\oldbibliography{#1}
\setlength{\baselineskip}{9.5pt}
\setlength{\itemsep}{2pt}} 

\title{Continuum limit of parton distribution functions from the pseudo-distribution approach on the lattice}
\ShortTitle{Continuum limit of PDFs from pseudo-distributions on the lattice}

\author*[a]{Manjunath Bhat}
\author[a]{Wojciech Chomicki}
\author[a]{Krzysztof~Cichy}
\author[b]{Martha~Constantinou}
\author[c]{Jeremy R.~Green}
\author[b]{Aurora~Scapellato}

\affiliation[a]{Faculty of Physics, Adam Mickiewicz University, ul.\ Uniwersytetu Pozna\'nskiego 2,\\ 61-614 Poznań, Poland}
\affiliation[b]{Temple University, 1925 N.\ 12th Street, Philadelphia, PA 19122-1801, USA}
\affiliation[c]{Deutsches Elektronen-Synchrotron DESY, Platanenallee 6, 15738 Zeuthen, Germany}
\emailAdd{manbha@amu.edu.pl}

\abstract{Precise exploration of the partonic structure of the nucleon is one of the most important aims of high-energy physics. In recent years, it has become possible to address this topic with first-principle lattice QCD investigations. In this talk, we focus on the so-called pseudo-distribution approach to determine the isovector unpolarized PDFs. In particular, we employ three lattice spacings to study discretization effects and extract the distributions in the continuum limit, at a pion mass of around 370 MeV. Also, for the first time with pseudo-PDFs, we explore effects of the 2-loop matching from pseudo- to light-cone distributions.}

\FullConference{%
The 39th International Symposium on Lattice Field Theory,\\
8th-13th August, 2022,\\
Rheinische Friedrich-Wilhelms-Universität Bonn, Bonn, Germany
}

\begin{document}
\maketitle
	
\section{Introduction}
\vspace*{-2mm}
Our knowledge of the universe has been influenced by studies of the structure of hadrons at high energy. One of the goals of particle physics is to determine how hadrons are built from their constituent particles. Despite being the primary component of visible matter and making up almost all of its mass, the nucleon has an internal structure that is still largely unexplored. Modern ongoing and planned experiments aim at providing vast amounts of data that will uncover several aspects of this structure. Ideally, this should be accompanied by first-principle investigations on the lattice. 
Over the years, the dominating thread on the lattice was to calculate differents kinds of form factors, i.e.\ moments of partonic distributions.
Despite early ideas from the 1990s and 2000s, the full $x$-dependence of these distributions could not be addressed, with the Euclidean spacetime metric being the main obstacle.
Almost a decade ago, Ji's concept to compute so-called quasi-distributions \cite{Ji:2013dva,Ji:2014gla} led to significant advances in this area.
While light-front correlations are inaccessible in Euclidean spacetime, Ji suggested that alternative observables that can be properly ``translated'' to the relevant distributions may be defined using lattice-calculable spatial correlations in a boosted hadron.
A number of lattice observables are suitable for deriving partonic distributions from them.
These observables must be renormalizable, have the same infrared structure as their light-front counterparts, and be calculable on the lattice.
Since Ji's initial idea, much theoretical and numerical effort has been devoted to understanding this approach and its alternatives, see e.g.\ the reviews~\cite{Cichy:2018mum,Ji:2020ect,Constantinou:2020pek,Cichy:2021lih,Cichy:2021ewm}. 

The same matrix elements that define quasi-distributions may also be utilized to construct another generalization of light-cone observables to Euclidean spacetime. Proposed by Radyushkin \cite{Radyushkin:2016hsy,Radyushkin:2017cyf,Radyushkin:2017lvu,Radyushkin:2018cvn,Radyushkin:2019mye}, this approach is known as pseudo-distributions.
Despite having the same underlying matrix elements, quasi- and pseudo-distributions evince significant differences.
The most essential one comes from their factorization into their light-front counterparts, performed in momentum space or in coordinate space, respectively.
Nevertheless, both approaches lead to the same physical distributions in the infinite momentum limit, upon subtraction of all lattice-specific and other systematic effects.

In this talk, we concentrate on two kinds of systematic effects, discretization and truncation effects. The former are addressed by doing our calculations at three lattice spacings and performing the continuum limit. For the latter, we implement for the first time the two-loop matching formulae \cite{Li:2020xml} and compare with one-loop effects~\cite{Radyushkin:2018cvn,Zhang:2018ggy,Izubuchi:2018srq}.
Full account of our work is given in the paper \cite{Bhat:2022zrw}.
For other work involving pseudo-PDFs, see Refs.~\cite{Orginos:2017kos,Karpie:2018zaz,Karpie:2019eiq,Joo:2019bzr,Joo:2020spy,Bhat:2020ktg,Karpie:2021pap,Egerer:2021ymv,HadStruc:2021wmh,HadStruc:2021qdf,HadStruc:2022yaw,Edwards:2022vtf}.

\vspace*{-3mm}
\section{Setup}
\vspace*{-3mm}
\subsection{Theoretical setup}
\vspace*{-2mm}
Nucleon quasi- and pseudo-PDFs, which are based on Euclidean correlations, are characterized by bare MEs, $\mathcal{M}_\Gamma(P,z)$, of the type $\left\langle P,s' \middle| \mathcal{O}_\Gamma(0,z) \middle| P,s \right\rangle = \mathcal{M}_\Gamma(P,z) \bar u(P,s') \Gamma u(P,s)$,
where $u(P,s)$ is a spinor corresponding to a Euclidean 4-momentum $P$ and spin $s$.
The bare non-local operator is $\mathcal{O}_\Gamma(x,z) \equiv \bar\psi(x) \Gamma \tau_3 W(x,x+z) \psi(x+z)$,
where $\vert P\rangle$ is a boosted nucleon state with four-momentum $P_\mu {=} (P_0, 0, 0, P_3)$ and $W(x,x+z)$ is a straight Wilson line. The Wilson line is chosen along the direction of the boost and has length $z$. The Ioffe time is given by $\nu=P_3z$.
Henceforth, we will specialize to flavor non-singlet ($u-d$) unpolarized distributions from $\Gamma=\gamma_0$ and drop the index $\Gamma$.
The $x$-dependent pseudo-PDF is given by
\vspace*{-2mm}
\begin{equation}
\label{eq:pPDF}
{\cal P} (x, z^2) = \frac{1}{2 \pi} \int_{-\infty}^\infty  d\nu \, e^{-i x \nu } \,  {\cal  M} (\nu, z^2).
\vspace*{-2mm}
\end{equation}

The Wilson line-induced power divergence and the conventional logarithmic divergence, which have been proven to be multiplicatively renormalizable to all orders in perturbation theory \cite{Ishikawa:2017faj,Ji:2017oey}, are both present in the matrix elements. These divergences may be eliminated by creating a double ratio using zero-momentum and local ($z=0$) matrix elements \cite{Orginos:2017kos}, which is referred to as reduced matrix elements or pseudo-ITDs (Ioffe time distributions),\vspace*{-2mm}
\begin{equation}
\label{eq:reduced}
\mathfrak{M}(\nu,z^2) = \frac{\mathcal{M}(\nu,z^2)\,/\,\mathcal{M}(\nu,0)}
{\mathcal{M}(0,z^2)\,/\,\mathcal{M}(0,0)}.
\end{equation}

\vspace*{-3mm}
\subsection{Matching to light-cone ITDs}
\vspace*{-2mm}
Through the use of a perturbative matching procedure, pseudo-ITDs may be matched to light-cone ITDs. We denote the matched ITDs by $Q(\nu,\mu)$ where $\mu$ is the renormalization scale. The relation between the PDFs $q(x, \mu)$ and the light-cone ITDs is given by\vspace*{-2mm}
\begin{equation}
\label{lightfr}
q(x,\mu)=\frac{1}{2\pi}\int^\infty_0 d\nu' e^{-i\nu' x}Q(\nu',\mu).
\vspace*{-2mm}
\end{equation}
One-loop perturbative evolution and matching formulae for the case of pseudo-ITDs were first derived in ~\cite{Radyushkin:2018cvn,Zhang:2018ggy,Izubuchi:2018srq} and the extension to two loops was given in \cite{Li:2020xml}. 
The two-loop matching formula is given by \vspace*{-2mm}
\begin{eqnarray}
\label{eq:inv}
Q(\nu,\mu;z)&=&\mathfrak{M}(\nu,z)-\frac{\alpha_s}{\pi} \int_0^1  du\, C^{(1)}(u)\left(\mathfrak{M}(u\nu,z)-\mathfrak{M}(\nu,z)\right)
-\frac{\alpha_s^2}{\pi^2} \int_{-1}^1  du\, C^{(2)}(u)\nonumber\\
&\times &\left(\mathfrak{M}(u\nu,z)-\mathfrak{M}(\nu,z)\right)
+\frac{\alpha_s^2}{\pi^2} \int_0^1  du\, C^{(1)}(u) \int_0^1  du'\, C^{(1)}(u')\nonumber\\
&\times &  \left(\mathfrak{M}(uu'\nu,z)-\mathfrak{M}(u\nu,z)-\mathfrak{M}(u'\nu,z)+\mathfrak{M}(\nu,z)\right),
\vspace*{-2mm}
\end{eqnarray}
where the form of $C^{(1)}(u)$ and $C^{(2)}(u)$ can be found in Ref.~\cite{Bhat:2022zrw}.

\vspace*{-3mm}
\subsection{Lattice setup}
\vspace*{-2mm}
We use lattice data from Ref.~\cite{Alexandrou:2020qtt}, where discretization effects were studied in the quasi-PDF approach.
For other Extended Twisted Mass Collaboration (ETMC) work involving quasi-distributions, we refer to Refs.~\cite{Alexandrou:2015rja,Alexandrou:2016jqi,Alexandrou:2017huk,Alexandrou:2018pbm,Alexandrou:2018eet,Alexandrou:2019lfo,Cichy:2019ebf,Green:2020xco,Chai:2020nxw,Bhattacharya:2020xlt,Bhattacharya:2020jfj,Alexandrou:2020zbe,Alexandrou:2020uyt,Bringewatt:2020ixn,Alexandrou:2021oih,Bhattacharya:2021moj,Alexandrou:2021bbo,Li:2021wvl,Bhattacharya:2021oyr,Bhattacharya:2022aob}.

Three ensembles of $N_f=2$ twisted mass fermions are used, produced by ETMC \cite{Baron:2010bv}, with lattice spacings $a=0.0644$ fm (ensemble D45), $0.0820$ fm (B55), and $0.0934$ fm (A60), at a pion mass of around 370 MeV.
The data of \cite{Alexandrou:2020qtt} at the source-sink separation of about $t_s=1$ fm and the boost $P_3\approx1.8$ GeV were supplemented with smaller-$P_3$ ones to explore the Ioffe-time dependence more completely and $P_3=0$ to determine the double ratios defining pseudo-ITDs.
For more details of the lattice setup, we refer to \cite{Alexandrou:2020qtt}.

\vspace*{-3mm}
\section{Results}
\vspace*{-2mm}
 \begin{figure}[t!]
\begin{center}
    \includegraphics[width=\textwidth]{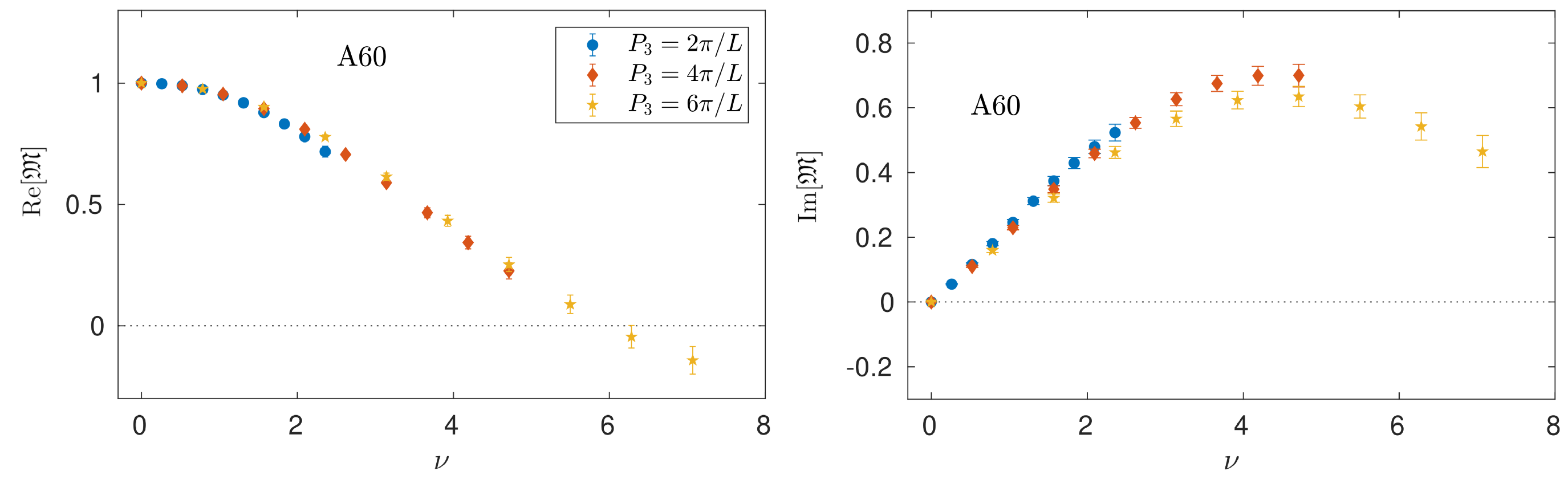}
    \includegraphics[width=\textwidth]{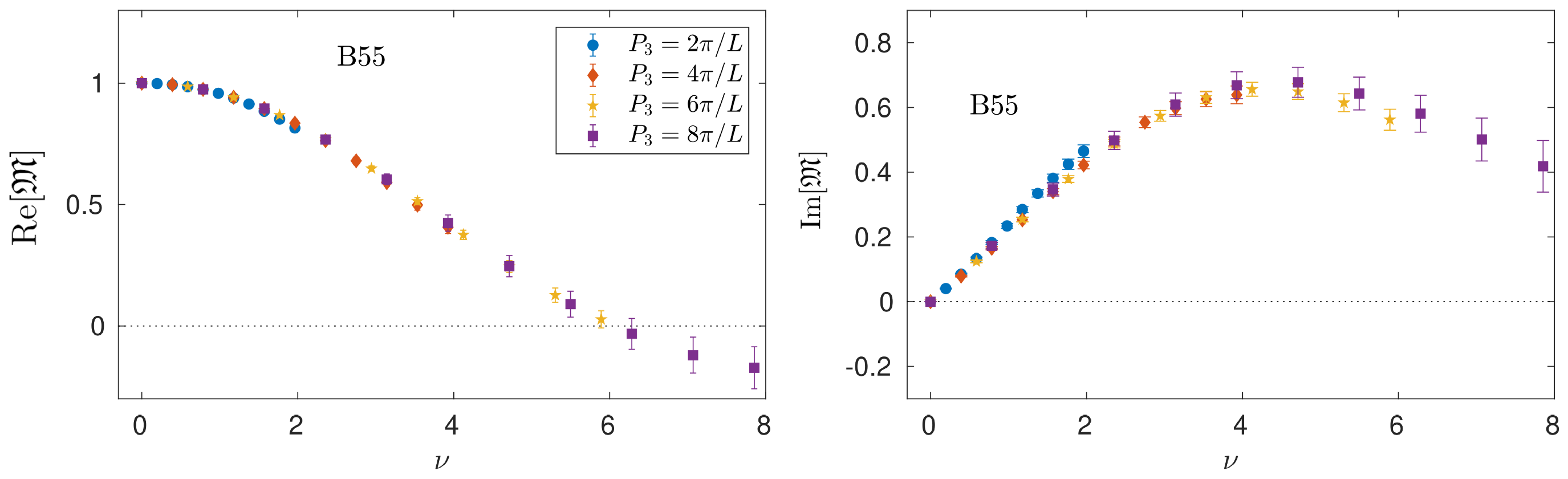}
    \includegraphics[width=\textwidth]{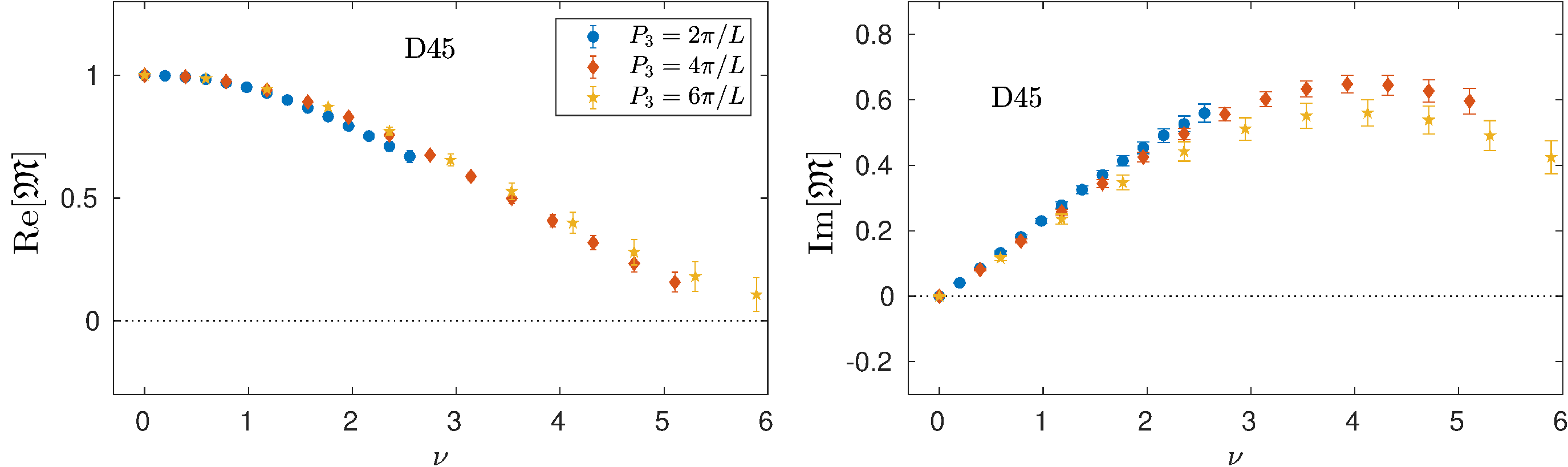}
\end{center}
\vspace*{-0.5cm} 
\caption{Real (left) and imaginary (right) part of reduced ITDs, $\mathfrak{M}(\nu,z)$, at different values of $P_3$ for the ensembles A60 (top), B55 (middle) and D45 (bottom).\vspace*{-2mm}}
\label{fig:reduced}
\end{figure}
We present the reduced ITDs as a function of the Ioffe time in Fig.~\ref{fig:reduced}. For all ensembles, the data from different boosts are close to lying on a universal curve. However, a trend can be observed that the real (imaginary) part increases (decreases) with the boost.

\begin{figure}[t!]
\begin{center}
    \includegraphics[width=\textwidth]{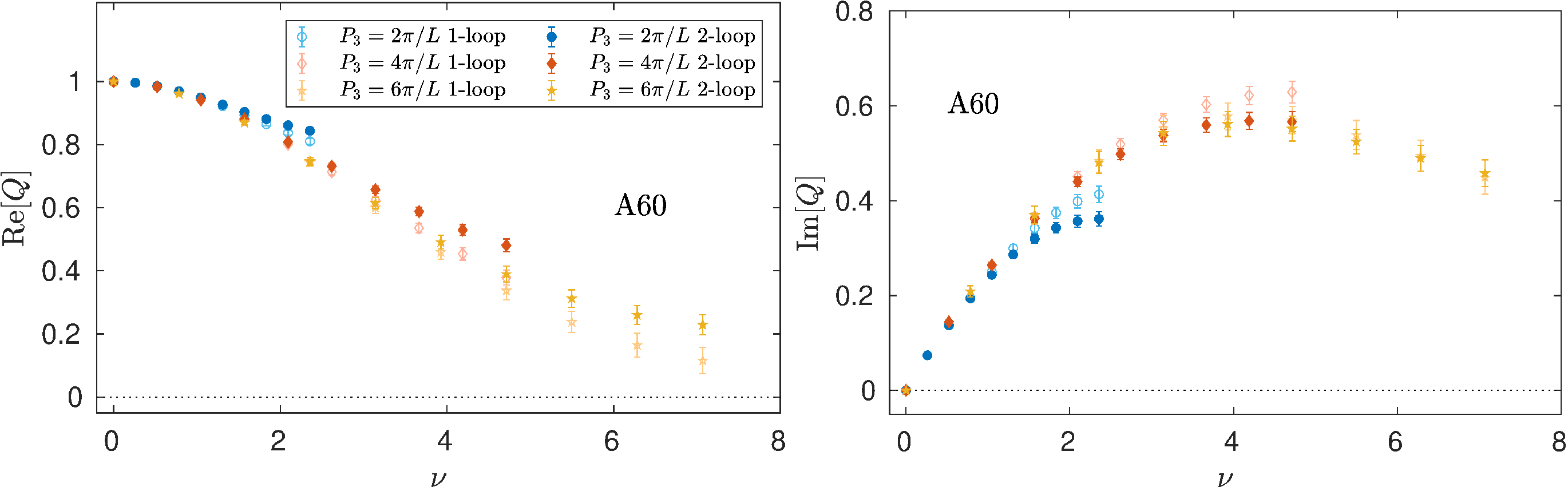}
    \includegraphics[width=\textwidth]{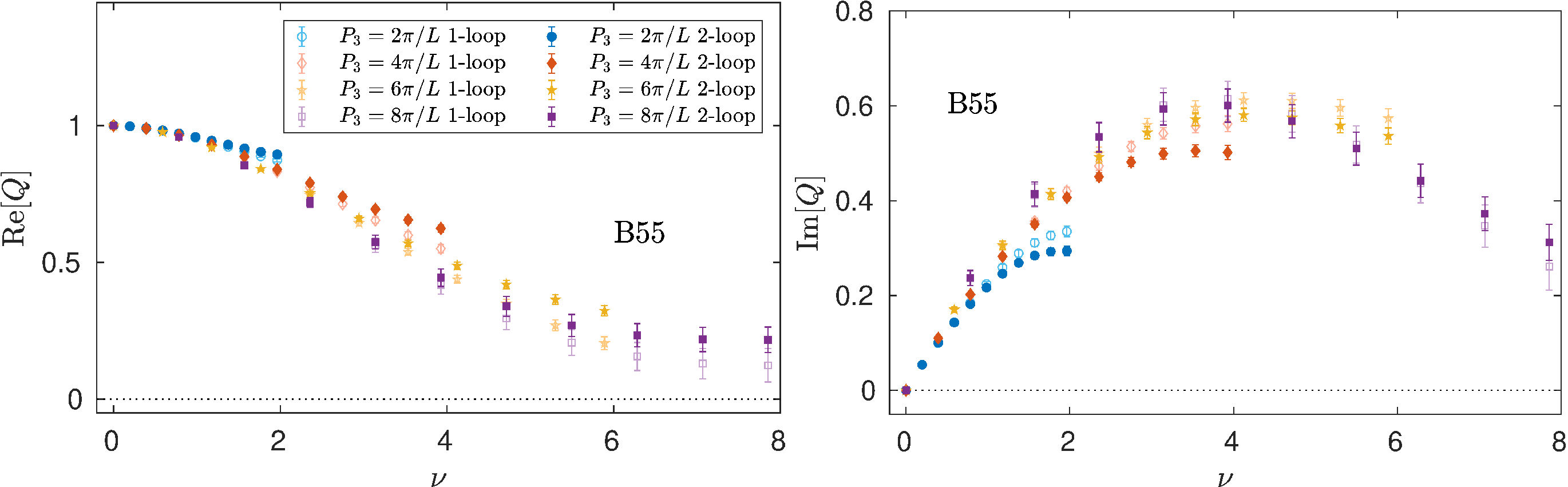}
    \includegraphics[width=\textwidth]{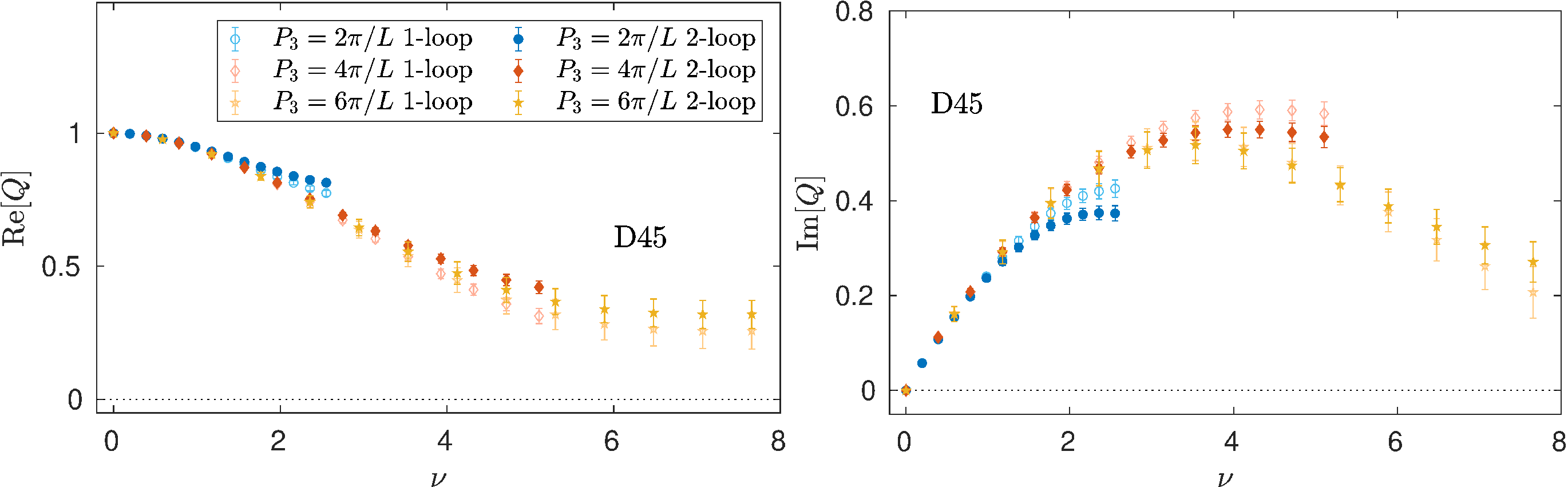}
\end{center}
\vspace*{-0.5cm} 
\caption{Real (left) and imaginary (right) part of matched ITDs, $Q(\nu,z)$, at different $P_3$ for the ensembles A60 (top), B55 (middle) and D45 (bottom) with one-loop (lighter colors) and two-loop (darker colors) matching.\vspace*{-2mm}}
\label{fig:matched}
\end{figure}
The matched ITDs at the level of various ensembles are shown in Fig.~\ref{fig:matched}. 
The perturbative evolution and matching is separated into its one-loop and two-loop parts.
Despite having a far smaller influence than the one-loop effect, the two-loop effect is becoming more important as Ioffe time increases.
Importantly, the perturbative procedure counteracts the trend observed for reduced ITDs and brings matched ITDs closer to the abovementioned universal curve, for small enough Ioffe times.
Clearly, at some point there is an overcompensation of the trend, leading to an opposite one of real (imaginary) part decreasing (increasing) with $P_3$.
This signals the breakdown of perturbation theory and in practice, it allows us to estimate values of $z$ that are consistent with short-distance factorization.
We adopt a criterion that the largest allowed $z\equiv\zmax$ is such that matched ITDs originating from different combinations of $(P_3,z)$, but fixed $P_3z$, are consistent within uncertainties.
As Fig.~\ref{fig:matched} suggests, this criterion leads to $\zmax^{\rm Re}\approx0.5$ fm and $\zmax^{\rm Im}\approx0.3$ fm.
Since we are ultimately interested in the continuum-extrapolated distributions, we observe that we can extend the latter also to 0.5 fm, upon error inflation by the continuum extrapolation.

\begin{figure}[h!]
\begin{center}
    \includegraphics[width=\textwidth]{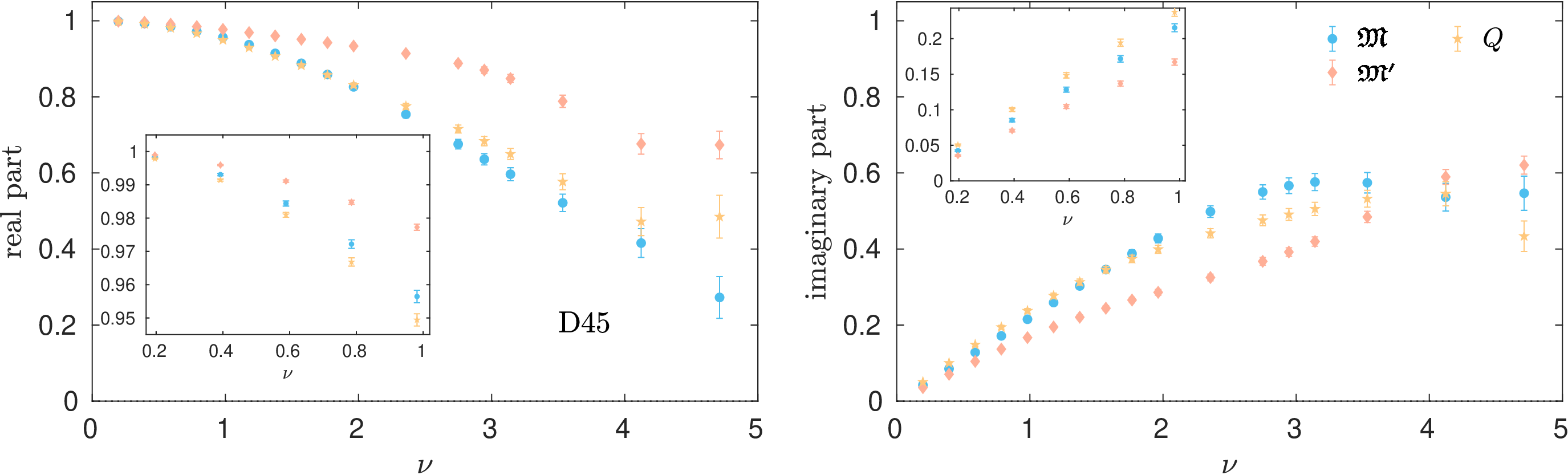}
\end{center}
\vspace*{-0.5cm} 
\caption{Real (left) and imaginary (right) part of reduced (blue circles), evolved (red rhombuses) and matched (yellow stars) ITDs, with the same Ioffe times corresponding to different combinations $(P_3,z)$ averaged over, including Wilson line lengths $z\leq\zmax=0.5$ fm for D45.\vspace*{-2mm}}
\label{fig:ITDs}
\end{figure}

All of the data that correspond to $z$ values that are greater than the value specified for $\zmax$ are discarded, and the ITDs that correspond to the same Ioffe times but are derived from different pairs of $(P_3,z)$ are averaged. 
A comparison of such $\nu$-averaged reduced, evolved, and matched ITDs (with two-loop equations) for the ensemble D45 is presented in Fig.~\ref{fig:ITDs} for $\zmax=0.5$ fm. 
We note that the two parts of the perturbative kernel, the evolution and matching/scheme conversion part, have opposite and almost equal in magnitude effects.
Thus, their net effect is small and matched ITDs are close to reduced ITDs, slightly above them for Re/small-$\nu$ and Im/large-$\nu$, and slightly below for Re/large-$\nu$ and Im/small-$\nu$.

\begin{figure}[t!]
\begin{center}
   \includegraphics[width=\textwidth]{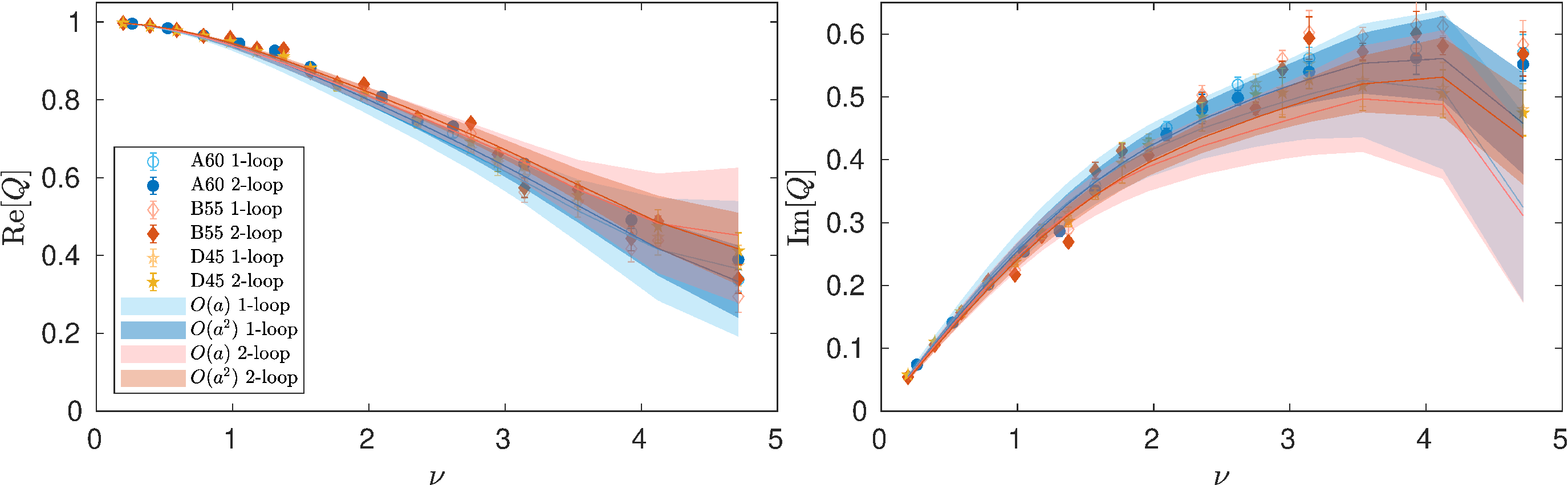}    
\end{center}
\vspace*{-0.5cm} 
\caption{Real (left) and imaginary (right) part of matched ITDs, where we averaged ITDs that have same the Ioffe time, for  $z\leq\zmax=0.5$ fm. Data points for ensembles A60 (blue circles), B55 (red rhombuses), and D45 (yellow stars) are shown. The bands correspond to matching peformed at one-loop/two-loop order, with the lighter/darker color pertaining to $\mathcal{O}(a)$/$\mathcal{O}(a^2)$ extrapolation to the continuum.\vspace*{-2mm}}
\label{fig:matched_nuavg}
\end{figure}
We perform $\mathcal{O}(a)$ and $\mathcal{O}(a^2)$ continuum limit extrapolations using $\zmax=0.5$ fm and one-loop or two-loop matched ITDs, see Fig.~\ref{fig:matched_nuavg}. 
As hinted above, the continuum-extrapolated ITDs have significantly inflated errors, within which results are consistent regardless of the extrapolation and perturbation theory order.

\begin{figure}[t!]
\begin{center}
    \includegraphics[width=\textwidth]{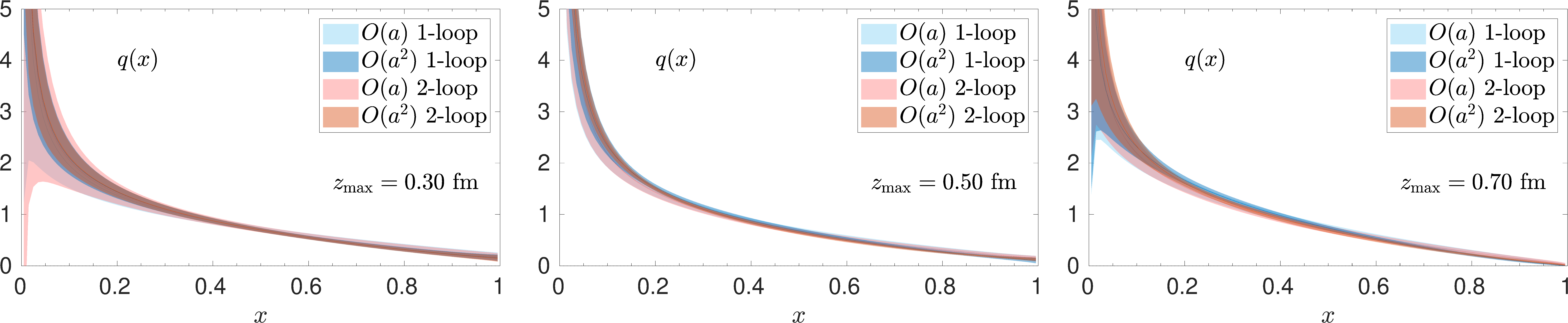}
    \includegraphics[width=\textwidth]{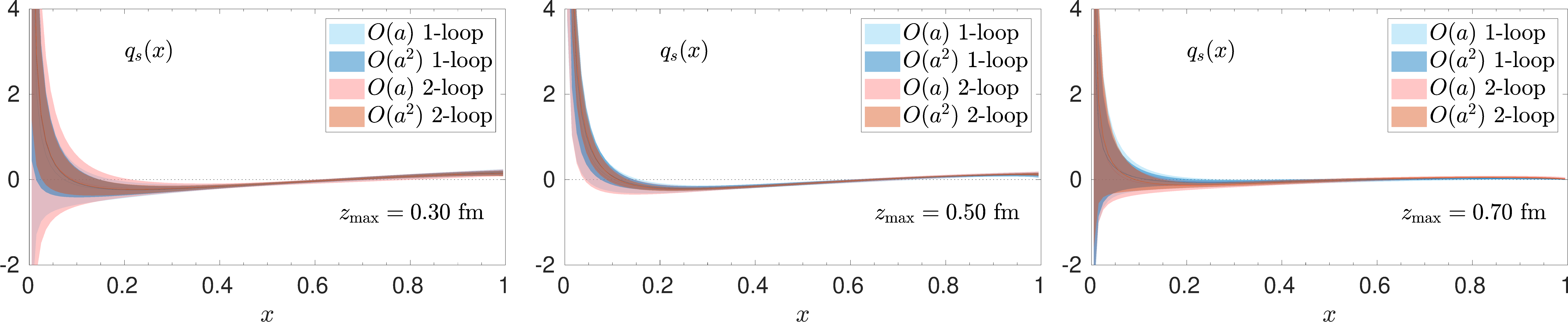}
\end{center}
\vspace*{-0.5cm} 
\caption{Continuum-extrapolated PDFs $q\equiv q_v+\bar{q}$ (top) and $\bar{q}$ (bottom) from the fitting ansatz reconstruction. From left to right: $\zmax=0.3,\,0.5,\,0.7$ fm. \vspace*{-2mm}}
\label{fig:cont}
\end{figure}

As the last step, we reconstruct the $x$-dependence using a fitting ansatz of the form $q_i(x) = N x^\alpha (1-x)^\beta$, where $i$ corresponds to the valence distribution $q_v$, related to the real part of ITDs, or the other non-singlet combination $q_v+2\bar{q}\equiv q_{v2s}$, related to the imaginary part.
Since $q_v$ is normalized to 1, only $\alpha$ and $\beta$ are fitting coefficients for the real part, while for the imaginary part, $N$ is an additional fitting parameter.
We also define the distributions $q\equiv q_v+\bar{q}=(q_v+q_{v2s})/2$ and $q_s\equiv\bar{q}=(q_{v2s}-q_v)/2$.
These two distributions are shown in Fig.~\ref{fig:cont}.
In this plot, we investigate also the influence of the choice of $\zmax$.
The results from $\zmax=0.3$ fm and $0.5$ fm are consistent with each other, while the ones from $\zmax=0.7$ fm show a qualitative difference in the large-$x$ regime.
Namely, both of the shown distributions at the two lower values of $\zmax$ are non-zero at $x=1$.
The behavior originates from the fits of the imaginary part and appears because even the larger value of $\zmax=0.5$ fm allows one to reach $\nu$ of only $4.7$ at $P_3\approx1.8$ GeV, where the ITD reaches its maximal value.
Thus, a significant part of the $\nu$-dependence is missing and in practice, a considerable subset of bootstrap samples in fits of Im$\,Q$ favors $\beta=0$ and thus, $q_{v2s}(x=1)\neq0$.
This bias is avoided with $\zmax=0.7$ fm, but as we argued above, such a large value of $\zmax$ renders the perturbative matching unreliable.
Hence, a robust reconstruction of the distributions involving the imaginary part ($q_{v2s}$, $q$ and $\bar{q}$) will only be possible if significantly larger Ioffe times can be reached at small enough $z$, i.e.\ with larger nucleon boosts.
We note that the problem does not appear in $q_v$, partially because of better stability of the fits due to their normalization condition (i.e.\ only two fitting parameters) and also due to the qualitatively different behavior of Re$\,Q$, monotonously decaying to zero with increasing $\nu$.

\begin{figure}[t!]
\begin{center}
    \includegraphics[width=\textwidth]{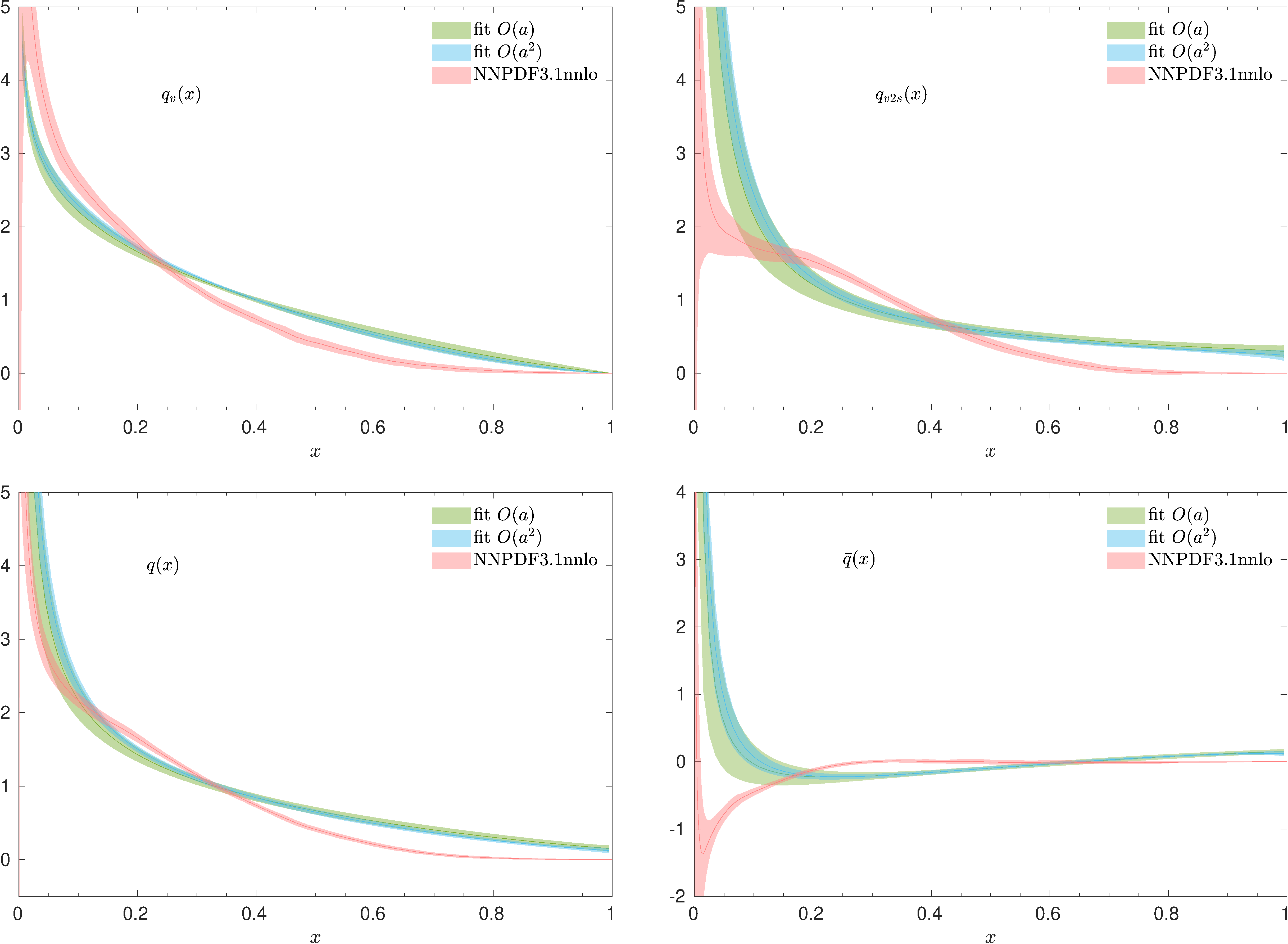} 
\end{center}
\vspace*{-0.5cm} 
\caption{Comparison of lattice-extracted PDFs with the corresponding NNPDFs (3.1, NNLO \cite{Ball:2017nwa}) for: $q_v$ (top left), $q_{v2s}=q_v+2\bar{q}$ (top right), $q=q_v+\bar{q}$ (bottom left), $q_s=\bar{q}$ (bottom right).\vspace*{-2mm} }
\label{fig:final}
\end{figure}

In Fig.~\ref{fig:final}, we present our final PDFs, $O(a)$-extrapolated to the continuum limit from two-loop matched ITDs with $\zmax=0.5$ fm, comparing them to the phenomenological extractions of NNPDF3.1 at NNLO \cite{Ball:2017nwa}. 
We observe that the statistical precision for $q_v$ is already similar to the precision of the valence NNPDF, even with the inflated errors due to the continuum limit extrapolation.
However, the uncertainty of the lattice result is not fully quantified.
Having taken the continuum limit, we eliminated one of the most natural sources of uncertainty, the one coming from cutoff effects.
This was computationally feasible at $m_\pi\approx370$ MeV, but the non-physical value of this mass introduces a possible bias.
The role of the non-physical pion mass is clear from the calculations of the first moment of $q_{v2s}$, $\langle x\rangle_{u-d}=\int_0^1 dx\, x(q_v(x)+2\bar{q}(x))$, found to be 40-70\% above the experimental value of $\langle x\rangle_{u-d}$ at $m_\pi\approx370$ MeV~\cite{Constantinou:2014tga}, e.g.\ $\langle x\rangle_{u-d}=0.270(3)$ for our ensemble B55, computed from local operators \cite{Abdel-Rehim:2015owa} (at the $t_s$ of our study).
The latter value is compatible with the ones we get by integrating the fitting-reconstructed $q_{v2s}$: $0.264(6)$ (A60), $0.254(5)$ (B55), $0.266(7)$ (D45), $0.269(25)$ ($\mathcal{O}(a)$ continuum limit of Fig.~\ref{fig:final} (top right)).
The enlarged $\langle x\rangle_{u-d}$ is manifested by enhancement of the PDF at $x\gtrsim0.5$.
This pion-mass-invoked behavior is seen also in $q_v$, the dominating input of $\langle x\rangle_{u-d}$.
Obviously, the enhanced value of $q_v$ at $x\gtrsim0.5$ implies its suppressed value in the small-$x$ region.
Analogous conclusions hold for $q=q_v+\bar{q}$ (bottom left panel of Fig.~\ref{fig:final}), where the most visible difference with respect to NNPDF occurs at large $x$.
Thus, we can speculate that the non-physical value of the light quark mass in our present work is the main systematic effect responsible for the difference between our PDFs and NNPDFs.
In turn, our current work excludes truncation effects in the matching as a significant effect.
Naturally, other sources of systematics need also be scrutinized. 

\noindent\textbf{Acknowledgments}.
  M.B., W.C.\ and K.C.\ acknowledge support by the National Science Centre
  (Poland) grant SONATA BIS no.\ 2016/22/E/ST2/00013. 
  M.C. acknowledges financial support by the U.S. Department of Energy, 
  Office of Nuclear Physics, Early Career Award under Grant No.\ DE-SC0020405. 
  J.R.G. acknowledges support from the Simons Foundation through the Simons Bridge for Postdoctoral Fellowships scheme.
  The calculations were performed at the Poznań Supercomputing and Networking Center (Eagle supercomputer) and at the J\"ulich Supercomputing Centre (JURECA \cite{jureca}), using the Grid library~\cite{Boyle:2016lbp} and the DD-$\alpha$AMG solver~\cite{Frommer:2013fsa} with twisted mass support~\cite{Alexandrou:2016izb}.

\bibliographystyle{h-physrev}
\vspace*{-4mm}
\bibliography{references.bib}
\end{document}